\documentclass[proceedings, preprint]{rmaa}

\usepackage{natbib}
\bibpunct{(}{)}{;}{a}{}{,}

\usepackage{paralist}

\usepackage{psfrag,color}

\SetYear{2009}
\SetConfTitle{Interferometric view of hot stars}

\title{Models of stars rotating near the critical limit} 

\author{
  G. Meynet,\altaffilmark{1} 
  C. Georgy,\altaffilmark{1}
  Y. Revaz\altaffilmark{2}
  R. Walder\altaffilmark{3}
  S. Ekstr\"om\altaffilmark{1}
  A. Maeder\altaffilmark{1}
  }

\altaffiltext{1}{Geneva Observatory, University of Geneva, 1290 Versoix, Switzerland (georges.meynet@unige.ch).}

\altaffiltext{2}{Laboratoire d'astrophysique de l'EPFL, 1290 Versoix, Switzerland (yves.revaz@epfl.ch).}

\altaffiltext{3}{Ecole Normale Sup\'erieure, Lyon, CRAL (UMR CNRS 5574), Universit\'e de Lyon, France,(rolf.walder@ens-lyon.fr).}

\shortauthor{Meynet et al.}
\shorttitle{RevMexAA}

\listofauthors{G. Meynet, C. Georgy, Y. Revaz, R. Walder, S. Ekstr\"om \& A. Maeder}
\indexauthor{Meynet, G.}
\indexauthor{Georgy, C.}
\indexauthor{Revaz, Y.}
\indexauthor{Walder, R.}
\indexauthor{Ekstr\"om, S.}
\indexauthor{Maeder, A.}

\abstract{When the surface angular velocity is above about 70\% of the critical angular velocity, many interesting features appear which may be tested by interferometric observations, like significant deformation of stars, variation of the effective temperature with the latitude. Also polar winds become important and equatorial disks may appear.  Near the critical limit, convection is also favored in the outer layers. In the present paper, we emphasize the need for a proper estimate of the critical velocity since this is the ratio of the actual velocity of the star to that critical velocity which determines the amplitude of the above effects. We recall the existence of two critical velocities. The first one, also called the classical critical velocity is the one to consider when the star has an Eddington factor inferior to 0.639, while the second one is the one to be considered when the Eddington factor is above 0.639.  The features  of the star at these two critical limits may  be very different.}

\addkeyword{Stars: Rotation}
\addkeyword{Stars: Early type}
\addkeyword{Stars: Mass loss}

\begin{document}
\maketitle

\section{The critical velocity}
\label{sec:intro}

The famous french writer Marcel Proust had the idea that the true discovery is not to discover new landscapes but to see them with new eyes.  This is probably what happens these days with interferometry which offers new eyes to observe the stars. For a very long time, the Sun was the only star for which the shape was observable. Interferometry now has obtained that information for a few stars as, {\it e.g.} Alta\"ir \citep{DeSouza2005, Monnier2007}, Vega \citep{Aufdenberg2005}, Regulus \citep{McAlister2005}, Achernar \citep{DeSouza2003,Carciofi2008}.

Among the very spectacular observation of this type is the very oblate shape obtained for the star Achernar (about 10 M$_\odot$). According to \citet{Carciofi2008}, the ratio of the equatorial to the polar radius is about 1.5, which corresponds to the expected ratio if two conditions are fullfilled: 1) the star rotates at the critical velocity, i.e. its linear velocity at the surface equator is equal to the Keplerian velocity there, 2) The Roche approximation used to compute the gravitational potential is valid. Let us recall that this approximation consists in using the gravitational potential due to a spherical distribution of mass for computing the gravitational force. This assumption is valid when only the outer layers of the star, which encompass only a tiny fraction of the total mass of the star, are heavily deformed. 

To check the first condition, it is necessary to compare the actual rotational velocity of the star to its critical one. Let us recall that the critical velocity, $\upsilon_{\rm crit}$,  is different from two other  often used velocities, the escape velocity, $\upsilon_{\rm esc}$,  and the Keplerian velocity $\upsilon_{\rm kep}$. One has that (in the frame of the Roche approximation)
$$\upsilon_{\rm crit}=\sqrt{G M \over R_{\rm ec}}=\sqrt{2 G M \over 3 R_{\rm pc}},$$
$$\upsilon_{\rm kep}=\sqrt{G M \over R_\mathrm{e}}\sim \sqrt{3 \over 2f}\upsilon_{\rm crit},$$
$$\upsilon_{\rm esc}=\sqrt{2G M \over R_\mathrm{e}}\sim \sqrt{6 \over f}\upsilon_{\rm crit}.$$
In the expressions above $G$ is the gravitational constant, $M$, the total mass of the star, $R_{\rm ec}$,  $R_{\rm pc}$ the equatorial, polar radius when the star is rotating at the critical velocity. The expression of $\upsilon_{\rm crit}$ as a function of $R_{\rm pc}$ makes use of the Roche approximation which implies that at the critical velocity the ratio of the equatorial to the polar radius is equal to 3/2. $R_{e}$ is the actual equatorial radius and $f=R_{e}/R_{p}$, with $R_{p}$, the actual polar radius. To estimate the Keplerian and escape velocity as a function of the critical one, we used the fact that the polar radius is weakly affected by rotation. Here we supposed a constant value irrespective of the rotation rate \citep[see Fig.~2 in][]{Ekstrom2008}.

We see from these expressions that using the expression of the Keplerian velocity, we overestimate the critical velocity by at most a factor $\sqrt{3/2}=1.225$\footnote{The Keplerian and the critical velocity are the same only if the star rotates at the critical velocity. This velocity  would also be  equal to the critical velocity of a star which  would not be deformed by the centrifugal acceleration.}. The escape velocity is at least (with $f=3/2$) a factor 2 and at most a factor 2.45 (with $f=1$) higher than the critical velocity.

\begin{figure}[!t]
\includegraphics[width=\columnwidth]{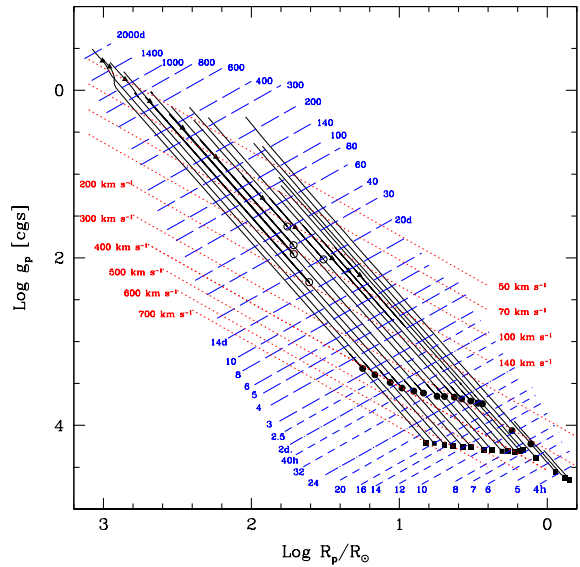}
\caption{Evolutionary tracks in the plane $\lg g_{\rm p}$ versus $\lg R_{\rm p}/R_\odot$, where  $ g_{\rm p}$ and $R_{\rm p}$ are the polar gravity and radius respectively. The black squares indicate the position of the ZAMS, the circles, the end of the MS phase, the triangles, the end of the core He-burning phase, the empty circles correspond to the position of the star at the blue end of a loop during the core He-burning phase. Superposed to the evolutionary tracks are indicated the critical periods (dashed lines) and critical equatorial velocities (dotted lines).  Only classical critical quantities are plotted (see text). The models used are those of \citet{Schaller1992} between 0.8 and 25 M$_\odot$.}
\label{fig:simple}
\end{figure}

The critical velocity is not a quantity easy to determine. The difficulties are the following:
\begin{itemize}
\item First the critical velocity has a different expression depending on the luminosity of the star \citep{Maeder2000}. If the Eddington factor $\Gamma= L \kappa /(4\pi c G M)$, where $\kappa$ is the total opacity, is higher than 0.639,  the critical velocity is no longer given by the expression above but by
$$\upsilon_{crit}={81 \over 16}{1-\Gamma_{\rm max}\over V^{'}_{\omega}}{R_{\rm e}^2(\omega)\over R^2_{\rm pc}},$$
where $R_{\rm e}$ is the equatorial radius for a given value of the rotation parameter $\omega$ which is the ratio of the actual angular velocity to the classical critical one, $\Gamma_{\rm max}$ the maximum Eddington factor over the surface (in general at the equator) and $V^{'}_{\omega}$ the ratio of the actual volume of the star to the volume of a sphere of radius $R_{\rm pc}$. In the following we shall focus on stars whith $\Gamma < 0.639$, so that the classical expression of the critical velocity can be used.
\item We can note that for computing the critical velocity we need to know the mass of the star and the equatorial radius it would have if it was rotating at the critical velocity. The actual star is not necessarily rotating at the critical velocity and thus its observed characteristics are not those the star would have if rotating at the critical limit. The critical velocity must therefore be obtained using some theoretical guide lines.
\end{itemize}
In case, it would be possible to obtain information on the polar radius as well as on the gravity at the pole\footnote{We use polar values because as mentionned above they are little affected by rotation. },
then the critical velocity can be obtained using the expression below:
$$\lg(\upsilon_{\rm crit})\sim 0.5\lg\left({R_{\rm p} \over R_\odot}\right)+0.5\lg(g_{\rm p})+0.5\lg\left({2 \over 3} R_\odot\right).$$
A similar expression can be obtained for the critical period.
$$\lg(P_{\rm crit})\sim\hfill\hfill$$
$$\hfill0.5\lg\left({R_{\rm p} \over R_\odot}\right)-0.5\lg(g_{\rm p})+0.5\lg\left({27 \over 2} \pi^2 R_\odot\right).$$
Figure~\ref{fig:simple} shows lines of constant $\upsilon_{\rm crit}$ and constant $P_{\rm crit}$ (critical period) in the plane $\lg g_{\rm p}$ versus $\lg R_{\rm p}/R_\odot$ where $g_{\rm p}$ is the gravity at the pole. 

\begin{figure*}[!t]
\includegraphics[width=\columnwidth]{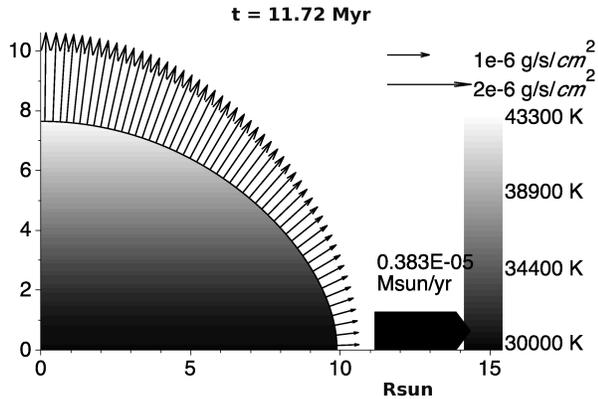}
\hspace*{\columnsep}
\includegraphics[width=\columnwidth]{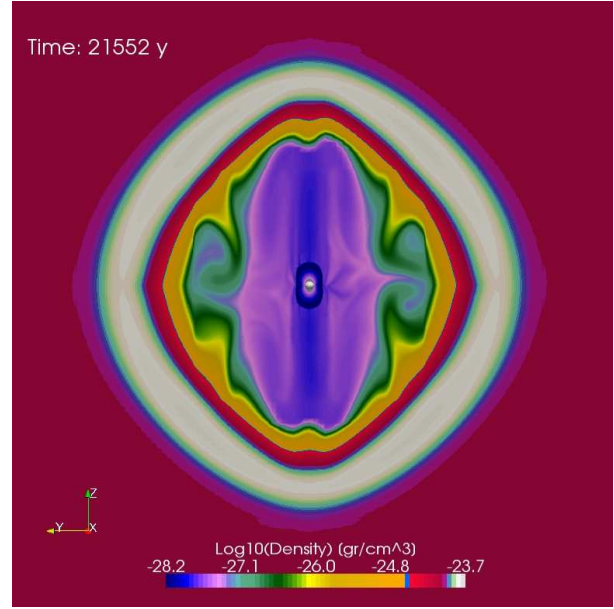}
\caption{{\it left:} Shape of a 20 M$_\odot$ star rotating near the critical limit at the end of the Main-Sequence phase. The variation of the effective temperature as a function of the latitude is indicated. One sees that the mass flux at the pole is more than 3 times the mass flux at the equator. There is also a mechanical mass loss at the equator. Its rate is about 4000 times the total mass loss due to line driven winds. {\it right:} The wind with its momentum and energy reshapes the gas of the circumstellar environment by pressing it into a nebula of much higher density than the ordinary circumstellar matter. The particular shape of the nebula will mirror the asymmetry of the wind itself. The case shown here corresponds to a 20 M$_\odot$ stellar model  having began its evolution with an equatorial rotation velocity on the ZAMS equal to 490 km s$^{-1}$ ($\Omega/\Omega_\mathrm{crit} = 0.75$). Are shown a slice in density normal to the equatorial plane of the star. The high density nebula (shown in white) is slightly aspherical. The low density cavity as well as the termination shock is highly aspherical. The wind material is basically confined in this low density cavity (shown in pink). However, up to 50 percent of material of the interstellar matter can be found in the outer layers of this cavity. The length of a side of the square containing the figure is 5 $\times$ 10$^{18}$ cm. }
\label{fig:widefig1}
\end{figure*}

\section{From the surface to the interior}

If indeed Archernar is rotating at the critical velocity, the fact that the ratio of equatorial to polar radius is equal to 1.5 can be interpreted as  a support to the Roche approximation. Now physically this approximation is correct as far as most of the mass of the star is not heavily deformed by rotation and this is indeed the case if most of the mass of the star rotates at velocities well below the local critical velocity. Thus in that case the shape of the surface tells us something about the way the interior is rotating. Models accounting for rotation following the prescription by \citet{Zahn1992} when they reach the critical velocity at the surface show an internal rotational profile far below the critical limit in qualitative agreement with the conditions required for the Roche approximation being valid. This results is not an effect of imposing the Roche approximation but of all the mechanisms affecting the angular momentum distribution in the star.

\section{What does happen at velocities near critical velocities?}

The situation is probably different in the case the critical velocity to be considered is the classical one $\upsilon_{\rm crit,1}$ or the one to be considered when the star has a near Eddington luminosity, {\it i.e.} $\upsilon_{\rm crit,2}$.

Let us first discuss a situation where the classical critical velocity has to be considered. In that case the following effects are expected:
\begin{itemize}
\item The combined effect of the line driven wind mechanisms and of the von Zeipel theorem induce some enhancement of the mass loss rate \citep[see Table~1 in][]{Maeder2000}. Let us note that while such a result is expected at first sight (higher mass loss because the gravity is on average reduced at the surface) a more careful analysis must also account for the fact that due to the von Zeipel theorem the radiative flux (which is the driving force) is also reduced!  Thus only a careful analysis can give the correct answer. We see that accounting for both effects produces an increase of the mass loss rates but an increase which remains modest due to the reduction of the radiative flux. 
\item When the dominating term in the opacity is electron scattering, then the wind will follow the variation of the radiative flux at the surface given by the von Zeipel theorem, namely the winds will be stronger in polar regions than in the equatorial ones. An illustration of this effect can be seen in Fig.~\ref{fig:widefig1} (left part). It can happen that when other opacity terms become dominant (typically when the star is in redder parts of the HR diagram than during the MS phase) the opacity may present strong variation with the latitude. The resulting wind will in that case be shaped by both the variation with the latitude of the radiative flux and that of the opacity. In addition to these effects, when the star is at the critical limit, then some mechanical mass loss will occur. Let us note that reaching the critical velocity does not mean that the mass will be ejected away from the star (the critical velocity is smaller than the escape velocity and e.g. Jupiter is orbiting around the Sun at the critical velocity but it is not escaping away), but the matter will be launched in a Keplerian orbit and an equatorial disk will be formed.
\item What are the consequences of wind anisotropies on the loss of angular momentum? We can first expect that less angular momentum will be lost (at least when the star is sufficiently hot for undergoing polar winds). However the situation is not so simple because while much less matter is lost at the equator, it is lost at a greater distance from the rotation axis!  Again here a careful estimate is needed for deducing a correct conclusion. It does appear that when careful account is made for the shape of the star and for the variation of the mass loss with the latitude, the wind anisotropies effects on the angular momentum of the star remains modest. 
\end{itemize}
What does happen when the second critical velocity is the one to be considered? The situation is quite different. One expect that
\begin{itemize}
\item the quantity of mass lost can be greatly enhanced due to the fact that the continuum and not only the line participate in driving the winds. The exact quantity of mass lost in that case remain to be determined. Obviously  the expressions used in the frame of the line driven winds can no long be used.
\item Because the deformation of the star is still given by the ratio $\upsilon/\upsilon_{\rm crit.1}$ and since $\upsilon_{\rm crit,2}$ can be much lower than $\upsilon_{\rm crit.1}$, the star in that case may not be heavily deformed and may also present nearly isotropic line driven winds. Interestingly, $\eta$Car presents polar winds. Thus, in the frame of the present model,  this implies that its surface velocity should not be too far from the classical critical velocity.
\end{itemize}

\bibliography{Biblio}


\end{document}